\documentclass[
  acmsmall,
  nonacm,
  screen
]{acmart}

\usepackage[T1]{fontenc}
\usepackage[utf8]{inputenc}

\usepackage{wrapfig}
\usepackage{adjustbox}
\usepackage{calc}

\usepackage{booktabs}
\usepackage{tabularx}
\usepackage{multirow}
\usepackage{multicol}

\usepackage{calc}

\usepackage[
  nolist
]{acronym}

\usepackage{siunitx}

\usepackage{color}
\usepackage[
  table
]{xcolor}
\usepackage{tcolorbox}

\usepackage{tikz}
\usetikzlibrary{
  positioning,
  calc,
  arrows,
  arrows.meta,
  shapes.misc,
  shapes.geometric,
  decorations.markings
}

\usepackage{tikz-layers}

\usepackage{tikz-layers}

\usepackage{adjustbox}

\usepackage{listings}

\usepackage{subcaption}
\usepackage{newfloat}

\DeclareFloatingEnvironment[
  fileext=lst,
  listname={List of Listings},
  name={List.},
  within=none
]{listing}

\DeclareCaptionSubType{listing}

\definecolor{mGreen}{rgb}{0,0.6,0}
\definecolor{mGray}{rgb}{0.5,0.5,0.5}
\definecolor{mPurple}{rgb}{0.58,0,0.82}

\newlength{\listingindent}
\lstdefinestyle{CStyle}{
  commentstyle=\color{mGreen},
  keywordstyle=\color{magenta},
  numberstyle=\tiny\color{mGray},
  stringstyle=\color{mPurple},
  basicstyle=\ttfamily\scriptsize,
  breakatwhitespace=false,
  breaklines=true,
  captionpos=b,
  keepspaces=true,
  numbers=none,
  numbersep=5pt,
  showspaces=false,
  showstringspaces=false,
  showtabs=false,
  tabsize=2,
  language=C
}

\newcommand{\code}[1]{\texttt{#1}}

\usepackage{xspace}

\usepackage{cleveref}

\usepackage{placeins}

\crefname{sublisting}{Listing}{Listings}
\Crefname{sublisting}{Listing}{Listings}

\makeatletter
\AtBeginDocument
 {
   \def\ltx@label#1{\cref@label{#1}}% add braces
   \def\label@in@display@noarg#1{\cref@old@label@in@display{#1}}% remove braces
\def\label@in@mmeasure@noarg#1{%
    \begingroup%
      \measuring@false%
      \cref@old@label@in@display{#1}% remove braces for multline, see https://tex.stackexchange.com/q/737204/2388
    \endgroup}%
 } %
\makeatother

\title{Checking the HAL Interface Specification Continuously, Right~from the Start}

\author{Manuel~Bentele}
\orcid{0009-0003-4794-958X}
\affiliation{
  \institution{University~of~Freiburg}
  \city{Freiburg}
  \country{Germany}
}
\affiliation{
  \institution{Hahn-Schickard Institute}
  \city{Villingen-Schwenningen}
  \country{Germany}
}
\author{Onur~Altinordu}
\orcid{0009-0005-2777-0763}
\affiliation{
  \institution{University~of~Freiburg}
  \city{Freiburg}
  \country{Germany}
}
\author{Jan~Körner}
\orcid{0009-0008-8631-9085}
\affiliation{
  \institution{University~of~Freiburg}
  \city{Freiburg}
  \country{Germany}
}
\author{Andreas~Podelski}
\orcid{0000-0003-2540-9489}
\affiliation{
  \institution{University~of~Freiburg}
  \city{Freiburg}
  \country{Germany}
}
\author{Axel~Sikora}
\orcid{0000-0003-0878-2919}
\affiliation{
  \institution{Offenburg University}
  \city{Offenburg}
  \country{Germany}
}
\affiliation{
  \institution{Hahn-Schickard Institute}
  \city{Villingen-Schwenningen}
  \country{Germany}
}

\makeatletter
\DeclareRobustCommand\onedot{\futurelet\@let@token\@onedot}
\newcommand*{\@onedot}{\ifx\@let@token.\else.\null\fi\xspace}
\makeatother

\newcommand{\eg}{e.g\onedot}

\newcommand{\Ie}{I.e\onedot}
\newcommand{\cf}{cf\onedot}

\newcommand{\wrt}{w.r.t\onedot}
\newcommand{\doublequote}[1]{``#1"}

\newcommand{\UltimateAutomizer}{\textsc{Ultimate Automizer}\xspace}
\newcommand{\STMCubeIde}{\textsc{STM32CubeIDE}\xspace}
\newcommand{\STMCubeMx}{\textsc{STM32CubeMX}\xspace}
\newcommand{\ModusToolbox}{\textsc{ModusToolbox}\xspace}
\newcommand{\DeviceConfigurator}{\textsc{Device Configurator}\xspace}
\newcommand{\Git}{\textsc{Git}\xspace}
\newcommand{\VSC}{\textsc{Visual Studio Code}\xspace}

\newcommand{\PSoC}{PSoC\xspace}
\newcommand{\STM}{STM32\xspace}
\newcommand{\SB}{SB17\xspace}
\newcommand{\Kimchi}{Kimchi\xspace}

\newcommand{\InfineonTechnologies}{Infineon Technologies\xspace}
\newcommand{\ST}{STMicroelectronics\xspace}

\newcommand{\opdep}{\triangleleft}
\newcommand{\varThadDep}[3]{#1 : #2 \opdep #3}

\begin{document}

  % !TeX spellcheck = en_US

\begin{acronym}
  \acro{adc}[ADC]{Analog-to-Digital Converter}
  \acro{hal}[HAL]{Hardware Abstraction Layer}
  \acro{idcc}[IDCC]{Incremental Development, Continuous Checking}
  \acro{cegar}[CEGAR]{Counterexample-guided Abstraction Refinement}
  \acro{posix}[POSIX]{Portable Operating System Interface}
  \acro{io}[I/O]{Input/Output}
  \acro{acsl}[ACSL]{ANSI/ISO C Specification Language}
  \acro{ide}[IDE]{Integrated Development Environment}
  \acro{cpu}[CPU]{Central Processing Unit}
  \acro{ram}[RAM]{Random-access Memory}
  \acro{dfg}[DFG]{Deutsche Forschungsgemeinschaft}
  \acro{spi}[SPI]{Serial Peripheral Interface}
  \acro{usb}[USB]{Universal Serial Bus}
  \acro{vcp}[VCP]{Virtual COM Port}
  \acro{uart}[UART]{Universal Asynchronous Receiver Transmitter}
  \acro{loc}[LOC]{Lines of Code}
  \acro{vcs}[VCS]{Version Control System}
  \acro{id}[ID]{Identifier}
  \acro{svcomp}[SV-COMP]{Competition on Software Verification}
  \acused{posix}
  \acused{io}
  \acused{acsl}
  \acused{ide}
  \acused{cpu}
  \acused{ram}
  \acused{dfg}
  \acused{spi}
  \acused{usb}
  \acused{uart}
  \acused{loc}
  \acused{id}
  \acused{svcomp}
\end{acronym}

  % !TeX spellcheck = en_US

\begin{abstract}
The correct use of a \ac{hal} interface in embedded applications is crucial to prevent malfunctions, crashes, or even hardware damage.
Software model checking has been successfully applied to check interface specifications in application programs, but
its employment in industrial practice is  hindered by its unpredictability (whether it succeeds for a given application program or not).
In this paper, we present a novel approach to address this problem by checking the \ac{hal} interface specification continuously and right from the start of the development.
\Ie, we develop an embedded application in several iterations without a formal connection between the steps.
The steps start from a \emph{program skeleton} which does nothing but calling \ac{hal} functions.
Actual functionality is added consecutively.
The \ac{hal} interface specification is checked in each step of the sequence.
The idea of the approach is to exploit a specific feature of software model checking: Its attempt to compute exactly the abstraction that is needed for the check to succeed may carry over from one step to the next, even if there is no formal connection between the steps.
The experience from a preliminary experimental evaluation of our approach in the development of embedded applications is very promising.
Following our approach, the check succeeds in each step and in particular in the final application program.
\end{abstract}

\keywords{Software Engineering, Incremental Development, Formal Verification, Program Analysis, Model Checking, Embedded Software, Temporal Properties, Hardware Abstraction Layer}

  \maketitle

  % !TeX spellcheck = en_US

\section{Introduction}
\label{sec:introduction}
When developing an application program for an embedded system that use a \acf{hal}, it is crucial to use the \ac{hal} interface correctly.
A misuse of the \ac{hal} interface may lead to serious issues including malfunctions, crashes, or even hardware damage.
For this reason, the specification of the \ac{hal} interface must be checked before the development of an application program is completed and the program is deployed.
Otherwise errors that violate the \ac{hal} interface specification remain undetected.
Software model checking has been successfully applied to check interface specifications in application programs, as several studies report, e.g., see~\cite{Ball2001,Ball2004,Kolb2010,Post2009}.
However, the employment of software model checking can be unpredictable.
It is not guaranteed that the check succeeds, especially if the application program becomes large and complex.

This paper addresses the challenge by presenting a novel development approach.
The approach includes an incremental workflow to develop application programs where the \ac{hal} interface specification is checked right from the start.
The incremental methodology drives the development of the application program in several iterations forward starting from an initial program with only calls of \ac{hal} functions.
In each step, the developer adds functionality to the application program.
Software model checking is then applied continuously to each revision of the program to detect violations of the \ac{hal} interface specification or to prove their absence.
To increase the chances of software model checking succeeds the development approach is structured in a way that the control flow of the program is first evolved, and then data flow is incorporated.

Unlike stepwise refinement approaches, \eg, see~\cite{Wirth1971,Back1990,Wehrheim2006,Ruhroth2012}, which establish a formal connection between program revisions, our approach is tailored for embedded software developers without requiring extensive expertise in formal methods.
Instead, we leverage software model checking which we apply independently to each program revision anew.
Here, we focus on checking the \ac{hal} interface specification at every development step.

The contribution of this paper is to present an approach for the incremental development of application programs that utilize a \ac{hal}.
For the checks of the \ac{hal} interface specification, our approach integrates software model checking into the application development.
This integration allows the application development to be adapted to software model checking in order to increase the chances that the checks succeeds.
We consider and present a specific \ac{hal} interface specification, which can be checked right from the start, even if there is no data flow within a program revision.
Based on an experimental evaluation, we demonstrate that the approach is feasible to develop two application programs, each for a different industrial embedded sensor system.

The paper is structured as follows:
In \Cref{sec:approach}, we introduce our overall approach to check the \ac{hal} interface specification during the development.
In \Cref{sec:incremental-development}, we provide insights on the incremental development of application programs, while \Cref{sec:continuous-checking} addresses the continuous checking.
\Cref{sec:evaluation} presents a preliminary experimental evaluation where we apply the approach to develop an application program for an industrial embedded sensor system.
Finally, \Cref{sec:related-work} offers an overview of related work before we conclude this paper.

  % !TeX spellcheck = en_US

\section{Overall approach \& workflow}
\label{sec:approach}
The development approach that we present in this paper integrates the checking of the \ac{hal} interface specification into an incremental workflow for application development.
This approach consists of two key elements:
the development of an application program through increments, and the continuous checking utilizing software model checking.
We refer to this development approach as \ac{idcc}.
The \emph{incremental} development adds functionality through several iterations, as opposed to the iterative development (\cf~\cite{Cockburn2008}).
The primary focus of \ac{idcc} is on checking the \ac{hal} interface specification right from the start of the application development.
The check is applied \emph{continuously} in each step of the development where it does not impose any formal connection to revisions of the program from previous steps.
Specifically, we focus on temporal properties for the correct \ac{hal} interface usage that can be checked at an early stage of development, even if there is no data flow within a program revision.

\begin{wrapfloat}{figure}{I}{17em}
  \centering
  \begin{subfigure}{8.25em}
    \begin{minipage}[b]{\linewidth}
      \centering
      % !TeX spellcheck = en_US

\begin{tikzpicture}[
    processstep/.style={draw,rectangle,rounded corners,minimum width=8em,minimum height=1.5em},
    processarrow/.style={draw,arrows={-Latex[width=0.5em, length=0.5em]}},
    node distance=0.9em and 3em,
    font=\small
  ]
  \node[processstep,fill=orange!50]             (n1) {Requirements};
  \node[processstep,below=of n1,fill=yellow!50] (n2) {Design};
  \node[processstep,below=of n2,fill=green!50]  (n3) {Implementation};
  \node[processstep,below=of n3,fill=cyan!50]   (n4) {Verification};
  \node[processstep,below=of n4,fill=red!50]    (n5) {Deployment};
  \node[processstep,below=of n5,fill=violet!30] (n6) {Maintenance};

  \path[processarrow] (n1) -- (n2);
  \path[processarrow] (n2) -- (n3);
  \path[processarrow] (n3) -- (n4);
  \path[processarrow] (n4) -- (n5);
  \path[processarrow] (n5) -- (n6);
\end{tikzpicture}
      \caption{Traditional.}%
      \label{fig:traditional-process-model}
    \end{minipage}
  \end{subfigure}
  \begin{subfigure}{8.25em}
    \begin{minipage}[b]{\linewidth}
      \centering
      % !TeX spellcheck = en_US

\begin{tikzpicture}[
    processstep/.style={draw,rectangle,rounded corners,minimum width=8em,minimum height=1.5em,align=center},
    processarrow/.style={draw,arrows={-Latex[width=0.5em, length=0.5em]}},
    node distance=0.9em and 3em,
    font=\small
  ]
  \node[processstep,fill=orange!50]             (n1) {Requirements};
  %\node[processstep,below=of n1,fill=gray!10,minimum height=4.5em+1.9em,execute at begin node=\setlength{\baselineskip}{1.9em}] (n2) {Design\\Implementation\\Verification};
  \node[processstep,below=of n1,fill=gray!10,minimum height=4.5em+1.85em] (n2) {};
    \node[processstep,below=of n1,minimum width=6.5em,yshift=-0.7em,execute at begin node=\setlength{\baselineskip}{1.25em},shading=axis,top color=yellow!50,bottom color=green!50] (n21) {Design\\Implementation};
    \node[processstep,below=of n1,fill=cyan!50,minimum width=6.5em,yshift=-4.25em] (n22) {Verification};
  \node[processstep,below=of n2,fill=red!50]    (n5) {Deployment};
  \node[processstep,below=of n5,fill=violet!30] (n6) {Maintenance};

  \path[processarrow] (n1) -- (n21);
  \path[processarrow] ($(n21.south)+(1.25em,0pt)$) to [bend left=15] ($(n22.north)+(1em,0pt)$);
  \path[processarrow] ($(n22.north)-(1.25em,0pt)$) to [bend left=15] ($(n21.south)-(1em,0pt)$);
  \path[processarrow] (n22) -- (n5);
  \path[processarrow] (n5) -- (n6);
\end{tikzpicture}
      \caption{\acs{idcc}.}
      \label{fig:idcc-process-model}
    \end{minipage}
  \end{subfigure}
  \caption{Traditional waterfall model~(\subref{fig:traditional-process-model}) compared to the \ac{idcc} approach as a subordinate workflow~(\subref{fig:idcc-process-model}) in the application development.}
  \label{fig:process-model}
\end{wrapfloat}
Traditional processes for software development, such as the waterfall model~\cite{Benington1983,Royce1987} illustrated in \Cref{fig:traditional-process-model}, typically consider software development and quality assurance separately in different process steps.
\ac{idcc} dissolves this separation by establishing a new workflow.
The workflow of \ac{idcc} imposes specific structural considerations on the design of program revisions.
These considerations are crucial for ensuring that each created program revision is beneficial for the continuous checking to succeed.
The approach of \ac{idcc} is flexible enough to be integrated into various existing development models, including traditional models such as the waterfall model~\cite{Benington1983,Royce1987} and agile-based practices like Scrum~\cite{Schwaber2020}.
It serves as a subordinate workflow within the software development to combine the traditional stages of \emph{Development}, encompassing \emph{Design} and \emph{Implementation}, with the \emph{Verification} stage, as shown in \Cref{fig:idcc-process-model}.

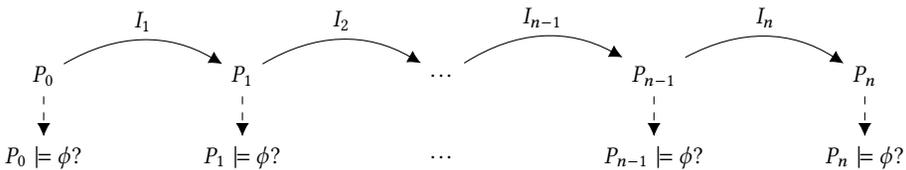
\begin{figure}[b]
  \centering
  % !TeX spellcheck = en_US

\begin{tikzpicture}[
    box/.style={minimum height=1.5em},
    processarrow/.style={draw,arrows={-Latex[width=0.5em, length=0.5em]}},
    node distance=1.5em and 6em,
    font=\small
  ]
  \node[box]             (p0) {$P_{0}$};
  \node[box,right=of p0] (p1) {$P_{1}$};
  \node[box,right=of p1] (p2) {\dots};
  \node[box,right=of p2] (p3) {$P_{n-1}$};
  \node[box,right=of p3] (p4) {$P_{n}$};

  \path[processarrow] (p0) edge[bend left] node[above] {$I_{1}$}   (p1);
  \path[processarrow] (p1) edge[bend left] node[above] {$I_{2}$}   (p2);
  \path[processarrow] (p2) edge[bend left] node[above] {$I_{n-1}$} (p3);
  \path[processarrow] (p3) edge[bend left] node[above] {$I_{n}$}   (p4);

  \node[box,below=of p0] (c0) {$P_{0} \models \phi$?};
  \node[box,below=of p1] (c1) {$P_{1} \models \phi$?};
  \node[box,below=of p2] (c2) {\dots};
  \node[box,below=of p3] (c3) {$P_{n-1} \models \phi$?};
  \node[box,below=of p4] (c4) {$P_{n} \models \phi$?};

  \path[processarrow,dashed] (p0) -- (c0);
  \path[processarrow,dashed] (p1) -- (c1);
  \path[processarrow,dashed] (p3) -- (c3);
  \path[processarrow,dashed] (p4) -- (c4);
\end{tikzpicture}
  \caption{Workflow of \ac{idcc} to develop an application program~$P_{n}$ step by step with $n$~increments $I_{1}, \dots, I_{n}$ while checking that each revision of the application program~$P_{i}$ of $P_{n}$ satisfies the interface specification~$\phi$.}
  \label{fig:idcc-workflow}
\end{figure}
The development workflow of \ac{idcc} begins with the given \ac{hal} interface specification and an initial program.
The program is developed by creating a sequence of increments through changes to the program's code, with each increment resulting in a new program revision.
Each created program revision is checked by applying software model checking in order to detect violations of the \ac{hal} interface specification.
If a program revision violates the \ac{hal} interface specification the bug causing the violation of the \ac{hal} interface specification has to be fixed first before a new increment is created.
Following the workflow visualized in \Cref{fig:idcc-workflow}, the aim is to arrive at a \emph{final program} that not only provides the required functionality but satisfies the \ac{hal} interface specification.
The final program can then be deployed without any misuse of the \ac{hal} interface.

One of the key benefits of \ac{idcc} is that it allows an early detection of specification violations in the development.
This benefit significantly reduces development effort and costs, as it enables immediate correction of faulty changes before they propagate further into the development progress.

  % !TeX spellcheck = en_US

\section{Incremental development}
\label{sec:incremental-development}
The development of a program revision takes place intermittent without any formal connection to a previous program revision.
\Ie, each program revision is a regular program provides some of the expected functionality of the final program.
The structure of a program revision is shaped by the initial program and the subsequent increments.
\ac{idcc} influences the development by guiding the creation of the initial program and the subsequent increments.
The basic idea of \ac{idcc} is to maximize control flow in early stages of the development.
By favoring control flow and gradual incorporation of additional data flow and computations with each increment, we can facilitate the checking task.
Program revisions with primarily control flow are often easier to check due to their reduced state space from fewer involved variables.

\subsection{Program Skeleton}
\label{sec:incremental-development-skeleton}
The initial program serves as \emph{program skeleton} from which development starts.
It lays the foundational structure for the creation of further increments.
A program skeleton is a blueprint to sketch the control flow of the \ac{hal} function calls for further program revisions, ideally for the final program.
The program skeleton is composed of \ac{hal} function calls without any variables employed.
The required \ac{hal} function calls are determined by the program's expected functionality.
When expressions within a program skeleton cannot be precisely determined we utilize non-deterministic choices, denoted by the symbol~\code{*}, or constant literals.
This is particularly important for arguments of the \ac{hal} function calls in the program skeleton.
In addition, we ignore return values of the \ac{hal} function calls and error handling.

\begin{wrapfloat}{listing}{I}{12em}
    \centering
    \fbox{%
        \begin{minipage}[t]{\linewidth-1em}
            % !TeX spellcheck = en_US

\begin{lstlisting}[style=CStyle]
void main()
{
  HAL_Init();
  HAL_UART_Receive(*);
  HAL_SPI_Transmit(*);
}
\end{lstlisting}

        \end{minipage}%
    }
    \caption{Program skeleton that only contains \ac{hal} function calls.}
    \label{lst:program-skeleton-example}
\end{wrapfloat}

An example of a program skeleton is shown in \Cref{lst:program-skeleton-example}.
The program skeleton contains only calls of the \ac{hal} functions \code{HAL\_Init}, \code{HAL\_UART\_Receive}, and \code{HAL\_SPI\_Transmit}.
It has no variables or data flow and the arguments of the \ac{hal} function calls are approximated by non-deterministic choices.
In such a skeleton, we can already check whether the order of the \ac{hal} function calls satisfies the correct use of the \ac{hal}.
If the program skeleton violates the \ac{hal} interface specification, it can be easily repaired.
This would be a more difficult task if the program is already more complex and overwhelming from a developer's perspective.

\FloatBarrier

\subsection{Increments}
\label{sec:incremental-development-increments}
\begin{wrapfloat}{listing}{I}{15em}
    \centering
    \setlength\fboxsep{1pt}
    \fbox{%
        \begin{minipage}[t]{\linewidth-0.25em}
            % !TeX spellcheck = en_US

\begin{lstlisting}[style=CStyle]
void main()
{
  <@\colorbox{green!20}{unsigned char data[7];}@>
  <@\colorbox{green!20}{struct message msg;}@>
  <@\colorbox{green!20}{int ret = 0;}@>

  <@\colorbox{green!20}{ret = }@>HAL_Init();
  <@\colorbox{green!20}{if (ret < 0) \{}@>
     <@\colorbox{green!20}{app\_error\_handler();}@>
  <@\colorbox{green!20}{\}}@>

  <@\colorbox{green!20}{while (1) \{}@>
    HAL_UART_Receive(<@\colorbox{green!20}{\&data}@>);
    <@\colorbox{green!20}{app\_deserialize(\&msg, \&data);}@>
    <@\colorbox{green!20}{if (msg.type == 0x1) \{}@>
      HAL_SPI_Transmit(<@\colorbox{green!20}{msg.cmd}@>);
    <@\colorbox{green!20}{\}}@>
  <@\colorbox{green!20}{\}}@>
}
\end{lstlisting}

        \end{minipage}%
    }
    \caption{Program revision which results from the changes of the increment colored in green.}
    \label{lst:increment-example}
\end{wrapfloat}
One or several changes of statements in the application program's code become an \emph{increment}.
Each change can either add, delete, or modify a statement in the current program revision.
If we apply an increment to a program revision~$P_{i}$, we obtain a new program revision~$P_{i+1}$.
In this way, increments allows us to refine the program skeleton in each development step in order to implement more functionality.
The development of increments continues until all desired functionality is implemented.
At the end of the development, we obtain the final program~$P_{n}$.
Each increment introduces increasing complexity by incorporating control flow and data flow while reducing non-deterministic choices by making expressions of statements more concrete.
Additionally, each increment records the code changes made during development, resulting in a comprehensive development history that serves as valuable documentation.

Lets revisit the example program skeleton from \Cref{lst:program-skeleton-example}.
We can add more functionality to the program skeleton by creating the increment shown in \Cref{lst:increment-example}.
The changes of the increment are colored in green and result in the shown program revision.
In this example, the increment extends the control flow by adding application-specific function calls, branches, and a worker loop.
The application-specific function calls include the functions \code{app\_error\_handler} and \code{app\_deserialize}.
The branches are utilized to restrict the specific call of the \code{HAL\_SPI\_Transmit} function and implement an error handling based on the return value of the called \code{HAL\_Init} function.
The added worker loop at the end of the program ensures that calls to the \ac{hal} functions \code{HAL\_UART\_Receive} and \code{HAL\_SPI\_Transmit} are repeated.
The increment also introduces data flow by adding variables of different data types.
Those variables are used in the example to refine the arguments of the \ac{hal} function calls and the expressions of the branching conditions.

\FloatBarrier

\subsection{Development guideline}
\label{sec:incremental-development-guideline}
When creating an increment, a developer can make many changes at once.
For example, the functionality for the final program, as shown in \Cref{lst:increment-example}, is implemented in only one increment.
The opposite extreme case would be to create a separate increment for each change of a single program statement.
However, this is not practical, as the number of increments explodes when developing larger programs.
Such worst cases are not conducive for \ac{idcc} and particularly for the continuous checking.
For that reason, we will provide recommendations to guide the development of subsequent increments.

The methodology for creating increments is significantly influenced by continuous checking.
The central idea of \ac{idcc} is to guide the development of increments in a way that software model checking succeeds.
In order to get closer to the goal, the guideline for the development includes two objectives that are pursued consecutively in the development of increments.

The first objective focuses on evolving the control flow of the program across multiple increments.
This objective includes the addition of application-specific functions without implementation which are integrated into the program through calls.
Changes related to the control flow of the program are implemented by adding loops and branches to handle various scenarios such as error handling.
Importantly, these changes are done without introducing new variables.
Instead, non-deterministic choices are utilized for all types of expressions where variables are used (\eg, branching conditions).
We then obtain a control flow structure that closely sketches the one from the final program.
Since there are no variables introduced, this objective leads to a program with a small state space.
The small state space increases the chances that software model checking succeeds while we already can check the correct usage of the \ac{hal} functions in the sketched control flow of the program.

Once a complex control flow is established, the second objective involves the addition of data flow across multiple increments.
This objective focuses on the refinement of the program by the added data flow while preserving the established control flow structure.
The objective includes the introduction of variables and instances of complex data types.
Those variables and instances are utilized in expressions where they allow us to reduce the previously inserted non-deterministic choices.
Additionally, computations that require these variables are incorporated to enhance the functionality of the program.
When larger memory blocks are needed for implementing the functionality, arrays can also be integrated.
It is recommended to create separate increments if variables, instances of complex data types, or arrays are introduced.
This prevents increments from becoming too large and helps to simplify the error localization if the \ac{hal} interface specification is violated.
Most likely the source of the error is related to recent changes of the current increment in which the variables, instances, or arrays are involved.

  % !TeX spellcheck = en_US

\section{Continuous checking}
\label{sec:continuous-checking}
Every time an increment is created, the resulting program revision must be checked to detect violations of the \ac{hal} interface specification during the development.
The increment is only considered \emph{complete} if the check succeeds and the \ac{hal} interface specification is satisfied.
For checking, we utilize software model checking to ensure the correctness of each program revision.

\subsection{\acs{hal} interface specification}
\label{sec:continuous-checking-tds}
Applying the check is most effective when the \acs{hal} interface specification can be checked at an early stage, such as in the initial program skeleton.
Otherwise, the \acs{hal} interface specification may be trivially satisfied, offering little value to the program correctness during the development.
The focus of \ac{idcc} is on a temporal property that is related to the correct use of the \ac{hal} functions.

The temporal property defines a dependency between \ac{hal} functions if the call of a \ac{hal} function depends on the previously fulfilled functionality of another \ac{hal} function.
For example, the \code{HAL\_SPI\_Transmit} function called in the program from \Cref{lst:increment-example} can only be used to receive data via an \ac{spi} controller if the hardware has previously been initialized by calling the \code{HAL\_Init} function.
More formally, for a pair of \ac{hal} functions $f_1$ and $f_2$, the dependency~$\varThadDep{\delta}{f_1()}{f_2()}$ (\doublequote{$f_2$ depends on $f_1$}) expresses that a call of $f_1$ must precede a call of $f_2$.
We call a pair of \ac{hal} functions a \emph{temporal dependency}.
We will here not formalize the semantics of a temporal dependency since it is intuitively clear.
We kept the expressiveness of a temporal dependency weak by design such that it can be already checked in the initial program skeleton.
Temporal dependencies are more specific than generic correctness properties (\eg, for overflow checking) but less detailed than full functional specifications.
Such properties cannot be directly checked by standard build tools like compilers.
Instead, their validity is encoded as program annotation including regular program statements of the C~programming language~\cite{ISO2011C}.

\begin{wrapfloat}{listing}{I}{13em}
    \centering
    \setlength\fboxsep{1pt}
    \fbox{%
        \begin{minipage}[t]{\linewidth-0.25em}
            % !TeX spellcheck = en_US

\begin{lstlisting}[style=CStyle]
<@\colorbox{gray!20}{int state\_d = 0;}@>

int HAL_Init() {
    int ret = <@$\dots$@>;

    <@\colorbox{gray!20}{state\_d = 1;}@>
    return ret;
}

void HAL_SPI_Transmit(...) {
    <@\colorbox{gray!20}{assert(state\_d == 1);}@>
    <@$\dots$@>
}
\end{lstlisting}

        \end{minipage}%
    }
    \caption{Encoding the validity of the temporal dependency~$\varThadDep{\delta}{\code{HAL\_Init()}}{\code{HAL\_SPI\_Transmit()}}$ through the validity of an assertion in the gray colored annotation of the application program.}
    \label{lst:temporal-dependency-annotation}
\end{wrapfloat}

We present the program annotation schematically in \Cref{lst:temporal-dependency-annotation}.
The annotation consists of auxiliary program statements of two kinds: assignment statements, which are not allowed to update program variables other than auxiliary variables, and assert statements.
For each temporal dependency~$\varThadDep{\delta}{f_1()}{f_2()}$, we insert an assignment statement in the code for the implementation of function~$f_1$ and an assert statement in the function~$f_2$, in addition to the declaration of an auxiliary variable for the temporal dependency~$\delta$.
In each execution, the value of the auxiliary variable flags whether the call of the function $f_1$ has (or has not yet) taken place.
The annotation modifies the code for the implementation of the \ac{hal} functions to which the temporal dependencies refer.
\Ie, the annotation does not concern the application-specific code of the application program.
In this sense, the encoding of the correctness of the application program \wrt the \ac{hal} is independent of the particular application code.
Consequently, the effort to encode the validity of temporal dependencies has to be done only once for a \ac{hal}.
The annotated \ac{hal} can then be reused for subsequent application developments.

\FloatBarrier

\subsection{Verification harness}
\label{sec:continuous-checking-harness}
\begin{wrapfloat}{listing}[20]{R}{15.5em}
    \centering
    \setlength\fboxsep{1pt}
    \fbox{%
        \begin{minipage}[t]{\linewidth-0.25em}
            % !TeX spellcheck = en_US

\begin{lstlisting}[style=CStyle]
void main()
{
  unsigned char data[7];
  struct message msg;
  int ret = 0;

  ret = HAL_Init();
  if (ret < 0) {
     app_error_handler();
  }

  while (1) {
    HAL_UART_Receive(&data);
    app_deserialize(&msg, &data);
    <@\colorbox{orange!20}{msg.type = *;}@>
    if (msg.type == 0x1) {
      HAL_SPI_Transmit(msg.cmd);
    }
  }
}
\end{lstlisting}

        \end{minipage}%
    }
    \caption{Required verification harness colored in orange to check the temporal dependency~$\varThadDep{\delta}{\code{HAL\_Init()}}{\code{HAL\_SPI\_Transmit()}}$.}
    \label{lst:verification-harness-example}
\end{wrapfloat}
During the development, it may happen that functions have already been declared and calls to them are part of the application program but their functionality has not yet been implemented.
In such cases, we have to assume that a function can return any possible value from the domain of its return type and it can modify the memory pointed to during the call.
We consider that the function only modifies the explicitly pointed memory and does not cause undefined behavior according to the C language standard~\cite{ISO2011C}.
To model the behavior of the function for verification, we employ a \emph{verification harness} with assumptions that approximate the possible effects of the function.
These assumptions directly influence the accuracy and soundness of the verification results.
Stricter assumptions limit the possible behavior of the function, while weaker assumptions allow for a broader range of behavior.

When using \ac{idcc}, the verification harness can be constructed in two ways.
One way is to create a detailed version that exactly reflects the full behavior of the function, which is precise but complex and costly.
The other way is to make a simplified, on-the-fly abstraction that over-approximates the behavior in a sound way.
The abstraction can be refined in each development step to become more precise ensuring that the verification remains sound while getting closer to the real function behavior.
In practice, a software developer can create a harness on the fly with \ac{idcc} by performing reachability checks on (finite executions of) the application program utilizing software model checking.
The developer checks whether all \ac{hal} function calls are reachable on these executions.
If they are, then the harness is considered adequate and does not need further refinement.
However, if a \ac{hal} function call is not reachable, it indicates that not all temporal dependencies are being checked.
In this case, the developer must extend the harness to make those function calls reachable.

For example, if we consider the program revision from \Cref{lst:increment-example} the behavior of the application-specific function \code{app\_deserialize(message,\,data)} is not yet implemented.
Since there is no verification harness, the memory of the supposedly parsed message remains unchanged.
As a consequence, the condition \code{msg.type\,==\,0x1} evaluates to \code{false} since the message \code{type} is set to the implicit default value~\code{0x0} which differs from the constant~\code{0x1}.
The branch with the \code{HAL\_SPI\_Transmit(...)} function call is then not reachable.
The validity of the temporal dependency $\varThadDep{\delta}{\code{HAL\_Init()}}{\code{HAL\_SPI\_Transmit()}}$ is not checked, even if violated.
We solve this issue in \Cref{lst:verification-harness-example} by adding a verification harness that consists of an additional assignment statement which assigns an non-deterministic value to the message \code{type}.
The assignment statement weakens the overly strict default behavior and allows the \code{HAL\_SPI\_Transmit} call to become reachable.

\FloatBarrier

\subsection{Software model checking}
For every program revision, we expect software model checking (\eg, see~\cite{Heizmann2013a}) to succeed.
With \emph{succeed} we mean that the software model checking tool yields a result that is either \emph{correct} (program revision satisfies the \ac{hal} interface specification) or \emph{incorrect} (program revision violates the \ac{hal} interface specification).
If a program revision violates the specification, we use the returned failure path to localize the bug in the faulty revision.
We can then repair the faulty revision by fixing the bug in the current increment.
However, if the software model checking tool fails to return a result (\eg, it runs out of resources), we consider such a case to be \emph{unknown} where a developer cannot reason anything about the correctness of the program revision \wrt the \ac{hal} interface specification.

The \ac{idcc} approach exploit a specific feature of software model checking.
Its attempt to compute exactly the abstraction needed for the check to succeed may carry over from one step to the next even if there is no formal connection between the steps.
Instead of manually creating the required abstraction, we rely on software model checking as a black box to compute it for us.
This ability to obtain the necessary abstraction automates the verification of temporal dependencies and facilitates the continuous checking.

  % !TeX spellcheck = en_US

\section{Experimental evaluation}
\label{sec:evaluation}
We evaluate the \ac{idcc} approach within the context of application development for industrial embedded systems, in particular sensor systems.
Specifically, we develop two separate application programs, each for a different sensor system.
The application programs are created by different software developers.
Both developments serve as a case study to investigate the application of \ac{idcc} in practice.
Here, we aim to answer the following research questions:
\begin{itemize}
    \item[\textbf{RQ1}:] Is it in principle feasible to apply the \ac{idcc} approach in the development of an application program for an industrial embedded system?
    \item[\textbf{RQ2}:] Can the verification harness be built up during the development with the \ac{idcc} approach?
    \item[\textbf{RQ3}:] Does a developer using the \ac{idcc} approach need to manually pass information (\eg, invariants) to the software model checking tool to ensure that continuous checking succeeds?
\end{itemize}
We use \UltimateAutomizer~\cite{Heizmann2018} as the software model checking tool for the continuous checking.
This tool is open-source and publicly available.\footnote{\UltimateAutomizer is part of the \textsc{Ultimate} program analysis framework: \url{https://ultimate-pa.org}}
The strong performance of the tool has been demonstrated  in the annual \ac{svcomp}~\cite{Beyer2024a}.

\subsection{Development environment}
The development environment for this experimental evaluation comprises two key elements.
First, the industrial embedded sensor systems.
Second, the software tools that are utilized for the development of the two application programs.

\subsubsection{Embedded sensor systems}
\begin{figure}
    \centering
    \includegraphics[width=0.75\linewidth]{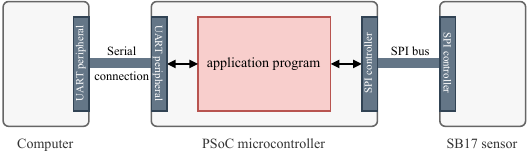}
    \caption{Architecture of the \SB sensor system for the development of the \PSoC application program.}
    \label{fig:sb17-embedded-system}
\end{figure}

The embedded sensor systems are similarly constructed.
Each of the two systems features a force sensor capable of measuring three-dimensional data based on bending moments.
Both systems consists of a microcontroller, the force sensor, and an external computer for data acquisition and processing.
The first system utilizes a \PSoC microcontroller from \InfineonTechnologies, connected to a \SB sensor chip.
This sensor chip is based on the design presented in~\cite{Kuhl2013}, with the original wireless interface for energy and data transfer replaced by an integrated \ac{spi} controller.
The architecture of this embedded sensor system is illustrated in \Cref{fig:sb17-embedded-system}.
The second system employs an \STM microcontroller from STMicroelectronics, combined with a \Kimchi sensor chip~\cite{Allinger2023}.
In both systems, the sensors are connected to their respective microcontrollers via the \ac{spi} bus.
The microcontrollers communicate with an external computer through \ac{uart} or \ac{usb} peripherals, respectively.
Each microcontroller contains a 32-bit ARM core that runs the application program, developed in the C programming language~\cite{ISO2011C}.

\subsubsection{Software development tools}
We use essential software tools for developing the two application programs.
For both developments, we manage code changes with \Git as our \ac{vcs}, which helps us keep track of code changes through structured commits.
To build the \SB sensor application for the \PSoC microcontroller, we use \VSC as \ac{ide}, along with \ModusToolbox, which is the official development ecosystem from \InfineonTechnologies.
Conversely, the \Kimchi sensor application for the \STM microcontroller is built using \STMCubeIde, which is the official \ac{ide} from \ST.
This tooling setup ensures that we evaluate the \ac{idcc} approach for developing application programs in a realistic manner.

\subsection{\ac{hal} interface specification}
For the application development, we utilize the official \ac{hal} for the \PSoC microcontroller,%
\footnote{\ac{hal} from \InfineonTechnologies for \PSoC microcontrollers: \url{https://infineon.github.io/mtb-hal-cat1/html/index.html}}
provided by \InfineonTechnologies, and the official \ac{hal} for the \STM microcontroller,%
\footnote{\ac{hal} from \ST for \STM microcontrollers: \url{https://github.com/STMicroelectronics/stm32f4xx-hal-driver}}
provided by \ST.
Both \acp{hal} provide a set of functions that can be called in the application program to access and operate hardware peripherals.
\begin{figure}
    \centering
    \adjustbox{max width=\linewidth}{%
        \centering
        % !TeX spellcheck = en_US

\begin{tikzpicture}
    \node (init) at (0,4) [rectangle,draw,font=\tiny,align=center,text width=1cm, rounded corners] {init};
    \node (free) at (-7.5,2) [rectangle,draw,font=\tiny,align=center,text width=1cm, rounded corners] {free};
    \node (freq) at (-6,2) [rectangle,draw,font=\tiny,align=center,text width=1cm, rounded corners] {set\_frequency};
    \node (ssc) at (-4.5,2) [rectangle,draw,font=\tiny,align=center,text width=1cm, rounded corners] {slave\_ select\_config};
    \node (sas) at (-3,2) [rectangle,draw,font=\tiny,align=center,text width=1cm, rounded corners] {select\_ active\_ssel};
    \node (recv) at (-1.5,2) [rectangle,draw,font=\tiny,align=center,text width=1cm, rounded corners] {recv};
    \node (send) at (0,2) [rectangle,draw,font=\tiny,align=center,text width=1cm, rounded corners] {send};
    \node (t) at (1.5,2) [rectangle,draw,font=\tiny,align=center,text width=1cm, rounded corners] {transfer};
    \node (ta) at (3,2) [rectangle,draw,font=\tiny,align=center,text width=1cm, rounded corners] {transfer\_ async};
    \node (busy) at (4.5,2) [rectangle,draw,font=\tiny,align=center,text width=1cm, rounded corners] {is\_busy};
    \node (abort) at (6,2) [rectangle,draw,font=\tiny,align=center,text width=1cm, rounded corners] {abort\_async};
    \node (regcall) at (3,6) [rectangle,draw,font=\tiny,align=center,text width=1cm, rounded corners] {register\_ callback};
    \node (event) at (6,6) [rectangle,draw,font=\tiny,align=center,text width=1cm, rounded corners] {enable\_event};

    \draw[-] (init) to[out=180,in=90] node[font=\tiny,left,pos=0.9 ]{$\delta_1$} node[fill=white!50,sloped,pos=0.5]{$\triangleright$} (free);
    \draw[-] (init) to[out=180,in=90] node[font=\tiny,left,pos=0.9 ]{$\delta_2$} node[fill=white!50,sloped,pos=0.5]{$\triangleright$} (freq);
    \draw[-] (init) to[out=180,in=90] node[font=\tiny,left,pos=0.9 ]{$\delta_3$} node[fill=white!50,sloped,pos=0.5]{$\triangleright$} (ssc);
    \draw[-] (init) to[out=180,in=90] node[font=\tiny,left,pos=0.9 ]{$\delta_4$} node[fill=white!50,sloped,pos=0.5]{$\triangleright$} (sas);
    \draw[-] (init) to[out=180,in=90] node[font=\tiny,left,pos=0.9 ]{$\delta_5$} node[fill=white!50,sloped,pos=0.5]{$\triangleright$} (recv);
    \draw[-] (init) to[out=270,in=90] node[font=\tiny,left,pos=0.9 ]{$\delta_6$} node[fill=white!50,sloped,pos=0.5]{$\triangleleft$} (send);
    \draw[-] (init) to[out=0,in=90] node[font=\tiny,left,pos=0.9 ]{$\delta_7$} node[fill=white!50,sloped,pos=0.5]{$\triangleleft$} (t);
    \draw[-] (init) to[out=0,in=90] node[font=\tiny,left,pos=0.9 ]{$\delta_8$} node[fill=white!50,sloped,pos=0.5]{$\triangleleft$} (ta);
    \draw[-] (init) to[out=0,in=90] node[font=\tiny,left,pos=0.9 ]{$\delta_9$} node[fill=white!50,sloped,pos=0.5]{$\triangleleft$} (busy);
    \draw[-] (init) to[out=0,in=90] node[font=\tiny,left,pos=0.9 ]{$\delta_{10}$} node[fill=white!50,sloped,pos=0.7]{$\triangleleft$} (abort);
    \draw[-] (init) to[out=0,in=180] node[font=\tiny,left,pos=0.9 ]{$\delta_{11}$} node[fill=white!50,sloped,pos=0.5]{$\triangleleft$} (regcall);
    \draw[-] (init) to[out=0,in=270] node[fill=white!50,sloped,pos=0.7]{$\triangleleft$} node[font=\tiny,right,pos=0.8]{$\delta_{12}$} (event);
    \draw[-] (regcall) to[out=0,in=180] node[font=\tiny,above,pos=0.8 ]{$\delta_{13}$} node[fill=white!50,sloped,pos=0.5]{$\triangleleft$} (event);
\end{tikzpicture}
    }
    \caption{Graph of temporal dependencies inferred from \ac{spi} functions of the \PSoC \ac{hal}.}
    \label{fig:psoc-hal-tds}
\end{figure}
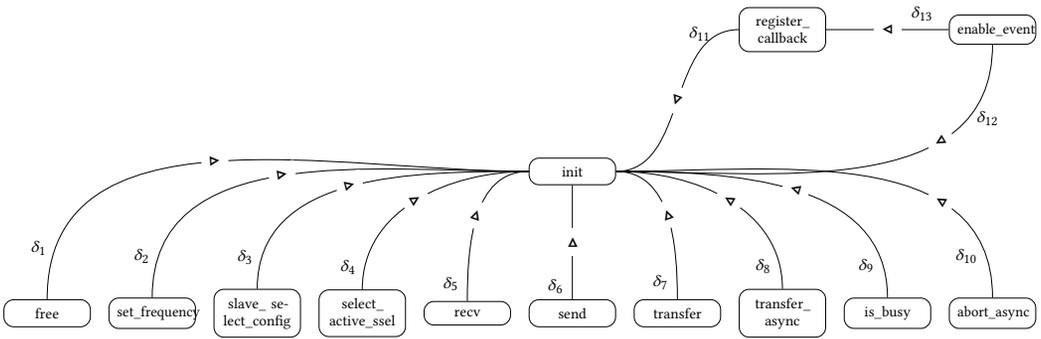
Before we start the application development, we consult the documentation and source code of the \acp{hal} to analyze and extract dependencies between the \ac{hal} functions.
In particular, we focus on the provided \ac{hal} functions to access and operate the \ac{spi} bus, as well as serial or \ac{usb} connections.
For the \PSoC \ac{hal}, we formulate a set of temporal dependencies where we obtained in total \num{33}~temporal dependencies (\num{13} from \ac{spi} functions and \num{20} from \ac{uart} functions).
Since the temporal dependencies constitute a strict partial order, we can represent them in an acyclic graph, as depicted with a subset of them in \Cref{fig:psoc-hal-tds}.
Similarly, for the \STM \ac{hal}, we formulate a set of temporal dependencies where we obtained in total \num{7}~temporal dependencies (\num{3} from \ac{spi} functions and \num{5} from \ac{usb} functions).
For continuous checking, we encoded the validity of the temporal dependencies as program annotation into the code of the corresponding \ac{hal} functions as described in \Cref{sec:continuous-checking-tds}.

\subsection{Experimental results}
In the following sections, we present the experimental results of our evaluation.
The two developers, who applied the \ac{idcc} approach, typically worked independently, with the exception of creating a shared communication library that implements a common protocol for transmitting measurements and receiving control commands from the external device.

\subsubsection{Final application programs}
The final application programs created by the developers implement an extensive functionality.
Both programs control their respective sensor chips and support all specific features that each sensor chip exposes via its \ac{spi} interconnect, such as the activation and deactivation of internal measurement units.
This allows for an optimized readout of sensor values, transferring only raw measurements from the activated units via the \ac{spi} bus.
The application programs preprocess the raw measurements to compute and store compensated measurement values.
These measurement values can be requested from the external computer via serial or \ac{usb} connection.
For the communication, we use messages defined by the Protocol Buffers specification%
\footnote{Protocol Buffers from Google: \url{https://protobuf.dev}}
to implement a communication library shared by both application programs.
The programs utilize serialization and deserialization functions generated by the Protocol Buffers compiler to handle communication messages.
We chose Nanopb\footnote{Nanopb from Petteri Aimonen: \url{https://jpa.kapsi.fi/nanopb}} for compilation because it is a lightweight version of Protocol Buffers optimized for embedded systems.
Both programs also support a bulk transfer mode.
In this mode, the measured values are sent automatically at set time intervals without requiring explicit requests from the external computer.

\begin{figure}
    \centering
    \includegraphics[width=0.6\linewidth]{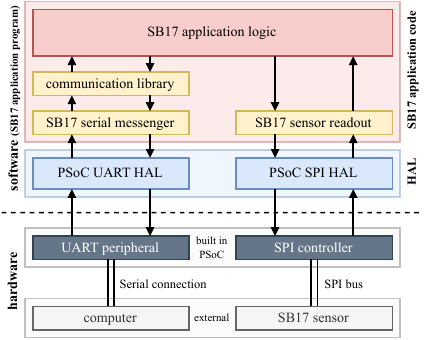}
    \caption{Architectural layers of the \PSoC application program for the SB17 sensor system.}
    \label{fig:sb17-application-layers}
\end{figure}

The application code of both programs contain complex control and data flow due to their extensive functionality.
The architecture of the final \PSoC program is shown in \Cref{fig:sb17-application-layers}.
The application code of the \PSoC program is split into several modules colored in yellow where each module provides a set of application-specific functions.
Those functions are used by the application logic to implement the behavior to control the embedded sensor system.
The arrows in the architecture indicate the data flow that has to be considered during the development to handle incoming requests and to read out and process measurements from the \SB sensor.

The final \STM program is structured similarly.
However, it uses a different \ac{hal}, specifically tailored for the \STM microcontroller, and a different peripheral device for the connection to the external computer.
Consequently, the application program implements an \ac{usb} messenger for the \ac{usb} peripheral instead of a serial messenger as depicted in \Cref{fig:sb17-application-layers}.
It also utilizes other \ac{hal} functions compared to the \PSoC application code.

\subsubsection{Program skeletons}
The application development begins with two initial programs that are generated either by the \DeviceConfigurator%
\footnote{\DeviceConfigurator from \InfineonTechnologies: \url{https://community.infineon.com/gfawx74859/attachments/gfawx74859/dexmc/269/1/Infineon-ModusToolbox_Device_Configurator_Guide_(Version_2.20)-Software-v01_00-EN.pdf}}
from the \ModusToolbox development ecosystem (see $P'$ in \Cref{tab:increments-sb17-development}) or by the \STMCubeMx generator%
\footnote{\STMCubeMx generator from \ST: \url{https://www.st.com/en/development-tools/stm32cubemx.html}}
from the \STMCubeIde.
The generated programs contain initialization code to set up hardware peripherals and provide a rough skeleton that allows the developers to insert application-specific program code.
However, the generated programs already incorporate data flow, which should not be present in skeletons that only call \ac{hal} functions.
Therefore, both developer decided to eliminate the data flow in the generated programs.
The developers were then able to insert the required \ac{hal} function calls for each of the two applications into the corresponding program, which result in two initial program skeletons (see $P_{0}$ in \Cref{tab:increments-sb17-development} for the \PSoC program skeleton).

\subsubsection{Increments}
Starting with the two initial program skeletons, several increments were created for each of the two application programs during development.

\begin{table}
    \centering
    \caption{Development history of the \PSoC application program for the embedded \SB sensor system.}%
    \label{tab:increments-sb17-development}
    \small
\newcommand{\hdr}[1]{\textbf{#1}}
\newcommand{\sdr}[1]{{\footnotesize\textbf{#1}}}
\newcommand{\unt}[1]{{\tiny\textbf{[#1]}}}
\newcommand{\increventry}[8]{#1 & #2 \newline {\scriptsize #3} & #4 & #5 & #6 & #7 & #8 \\}
\begin{tabularx}{\textwidth}{lXc|lr|rr}
    \toprule
    \multicolumn{3}{c|}{\hdr{Increment}}              & \multicolumn{2}{c|}{\hdr{Progam}}                                   & \multicolumn{2}{c}{\hdr{Checking}} \\
    \sdr{\ac{id}} & \sdr{description} & \sdr{extends} & \multicolumn{1}{c}{\sdr{rev.}} & \multicolumn{1}{c|}{\sdr{size}}    & \multicolumn{1}{c}{\sdr{time}} & \multicolumn{1}{c}{\sdr{mem.}} \\
                  &                   & \sdr{harness} & & \multicolumn{1}{c|}{\unt{\ac{loc}}} & \multicolumn{1}{c}{\unt{s}} & \multicolumn{1}{c}{\unt{GB}} \\
    \midrule
    \increventry{$I'$}{Add changes that result in the initial program.}
                {The initial program is generated by the \DeviceConfigurator of the \ModusToolbox development ecosystem from \InfineonTechnologies.}
                {---}{$P'$}{23}{22.97}{0.74}
                % Program revision size: 8 (blank) + 15 (code) = 23 (application code without comments + annotated HAL)
                % Program revision files: 1
    \increventry{$I_{0}$}{Add calls of required \ac{hal} functions to program.}
                {The changes comprise \ac{hal} function calls but no branches or loops are added. The application-specific error handling from $P'$ is removed to eliminate data flow for the initial program skeleton.}
                {\checkmark}{$P_{0}$}{615}{34.20}{1.22}
                % Program revision size: 199 (blank) + 416 (code) = 615 (application code without comments + annotated HAL)
                % Program revision files: 12
    \increventry{$I_{1}$}{Add application-specific functions and their calls.}
                {The application-specific functions are declared and are meant to implement the \ac{uart} communication handling. Their calls are inserted into $P_{0}$ but no branches or loops are added.}
                {---}{$P_{1}$}{1193}{38.31}{1.46}
                % Program revision size: 361 (blank) + 832 (code) = 1193 (application code without comments + annotated HAL)
                % Program revision files: 12
    \increventry{$I_{2}$}{Add branches and loops with non-deterministic choices.}
                {The branches and loops extend the control flow where their conditions are represented by non-deterministic choices. No variables are added.}
                {---}{$P_{2}$}{1407}{50.80}{1.34}
                % Program revision size: 373 (blank) + 1034 (code) = 1407 (application code without comments + annotated HAL)
                % Program revision files: 20
    \midrule
    \increventry{$I_{3}$}{Add error handling with required variables.}
                {The error handling checks the return value of called \ac{hal} functions and modifies the control flow of the program accordingly. The variables required for it are of a primitive data type.}
                {---}{$P_{3}$}{1494}{50.36}{1.50}
                % Program revision size: 375 (blank) + 1119 (code) = 1494 (application code without comments + annotated HAL)
                % Program revision files: 20
    \increventry{$I_{4}$}{Add variables of primitive data types for computations.}
                {The variables are used in added computations. The computations are utilized for reducing non-non-deterministic choices to make expressions more concrete.}
                {---}{$P_{4}$}{1602}{48.75}{1.50}
                % Program revision size: 420 (blank) + 1182 (code) = 1602 (application code without comments + annotated HAL)
                % Program revision files: 20
    \increventry{$I_{5}$}{Add instances of complex data types for computations.}
                {The data types of the instances are application-specific or provided by the \ac{hal}. The instances store sensor or \ac{hal} related data which is passed to or obtained from function calls.}
                {---}{$P_{5}$}{1676}{54.15}{1.54}
                % Program revision size: 439 (blank) + 1237 (code) = 1676 (application code without comments + annotated HAL)
                % Program revision files: 20
    \increventry{$I_{6}$}{Add instances of complex data types for communication.}
                {The data types of the instances are application-specific. The instances hold messages and states of the serial communication with the external computer.}
                {\checkmark}{$P_{6}$}{1766}{66.20}{1.48}
                % Program revision size: 465 (blank) + 1301 (code) = 1766 (application code without comments + annotated HAL)
                % Program revision files: 20
    \increventry{$I_{7}$}{Add array handling for initial \SB sensor check.}
                {Arrays are added to check the initial \ac{spi} communication through comparing an expected \ac{id} received from the \SB sensor.}
                {\checkmark}{$P_{7}$}{1775}{64.16}{1.44}
                % Program revision size: 468 (blank) + 1307 (code) = 1775 (application code without comments + annotated HAL)
                % Program revision files: 20
    \increventry{$I_{8}$}{Add array handling for reading out sensor measurements.}
                {Arrays are added to store data packets that are either sent or received via \ac{spi} bus from the \SB sensor. Another array holds the raw data of the \SB sensor measurements.}
                {\checkmark}{$P_{8}$}{1794}{71.43}{1.49}
                % Program revision size: 475 (blank) + 1319 (code) = 1794 (application code without comments + annotated HAL)
                % Program revision files: 20
    \increventry{$I_{9}$}{Add computation of compensated sensor measurements.}
                {The computation is encapsulated in an application-specific function and calculates the compensated \SB sensor measurements from the raw data.}
                {---}{$P_{9}$}{1810}{67.08}{1.76}
                % Program revision size: 479 (blank) + 1331 (code) = 1810 (application code without comments + annotated HAL)
                % Program revision files: 20
    \increventry{$I_{10}$}{Add array handling for readout configuration.}
                {The added array holds the configuration for the range-based readout of active sensor units in the \SB sensor chip.}
                {---}{$P_{10}$}{1825}{67.84}{1.54}
                % Program revision size: 483 (blank) + 1342 (code) = 1825 (application code without comments + annotated HAL)
                % Program revision files: 20
    \bottomrule
\end{tabularx}

\end{table}

The developer of the \PSoC application program created in total \num{10}~increments.
A detailed list of all these increments is provided in \Cref{tab:increments-sb17-development}.
In the first two increments ($I_{1}$ and $I_{2}$), the developer only added control flow to the \PSoC application program by declaring and calling application-specific functions while inserting branches and loops.
In the following increments, the developer introduced data flow by adding variables and instances of complex data types.
In the last three increments ($I_{7}$, $I_{8}$, and $I_{9}$), the developer allocated memory blocks for storing sensor measurements values by inserting arrays into the application code.

Similarly, the developer of the \STM application program created in total \num{9}~increments.
The development history looks similar to that of \PSoC development.
However, the developer of the \STM application program uses a different strategy to create the increments.
Instead of implementing a part of the functionality within a single increment, the developer has distributed it across multiple increments.
For example, the data flow for refining the \ac{hal} function calls for initializing the hardware peripherals is distributed across several increments.

\subsubsection{Verification harness}
During the creation of the increments for the \PSoC application program, the developer extended the verification harness (\cf \Cref{tab:increments-sb17-development}) by incorporating several assumptions.
When creating the initial program skeleton in increment~$I_{0}$, the developer added non-determinism to the C~program skeleton to simulate the behavior of interrupt calls.
The developer further added assumptions to model the behavior of a function pointer for a callback function.
The call of the callback function has to be included in the control flow of the initial program skeleton.
In the increments~$I_{6}$, $I_{7}$, and $I_{8}$, the developer added environmental assumptions that state the presence of structured data in memory blocks when calling \ac{hal} functions for receiving data from the \ac{spi} bus and the serial connection.
The added assumptions in $I_{6}$ also include the modeling of behavior for serialization and deserialization functions from the Nanopb implementation.
These functions are provided in a pre-compiled library, which is linked to the application code at the end of the development process.

The developer of the \STM application program followed a similar approach when creating the harness.
But there is a notable difference: Interestingly, the developer added non-determinism for the simulation of interrupt calls in a later increment, as the hardware peripherals that may trigger interrupts are initialized in a later increment.

Both developer successfully built their verification harnesses throughout the development workflow using the \ac{idcc} approach~(\textbf{RQ2}).

\subsubsection{Continuous checking}
The developers performed continuous checking at each program revision to detect violations of temporal dependencies.
The software model checking carried out by \UltimateAutomizer succeeds for each revision of the two application programs.
This demonstrates that the application development using the \ac{idcc} approach is feasible (\textbf{RQ1}).

The results for each check of the \PSoC application program are shown on the right in \Cref{tab:increments-sb17-development}.
The size of the resulting program revisions refers to the size of the created \PSoC application code and ranges from \num{23} to \num{1825}~lines of code (excluding comments).
\Ie, the size of each program revision does not contain the size of the annotated \ac{hal}.
The annotated \ac{spi} and \ac{uart} \ac{hal} involves \num{344} more lines of code (excluding comments), included in each check.
\UltimateAutomizer takes at most \qty{72}{\second} to check a \PSoC program revision while the memory consumption never exceeds \qty{1.8}{\giga\byte}.

The results for each check of the \STM application program are of the same order of magnitude as the results of the \PSoC application program.
The size of the \STM program revisions ranges from \num{677} to \num{1266}~lines of code (excluding comments).
\UltimateAutomizer takes at most \qty{214}{\second} to check a \STM program revision while the memory consumption never exceeds \qty{1.8}{\giga\byte}.

The checking results from both application developments indicate that software model checking by \UltimateAutomizer can be carried out on a regular desktop computer within a time that is acceptable for a developer.\footnote{All verification runs with \UltimateAutomizer in version 0.3.0-dev-11019c6 were carried out and measured on a regular desktop computer with quad-core \ac{cpu} at \qty{3.4}{\giga\hertz}, \qty{8}{\giga\byte} of memory.}
This is particularly relevant for developers who apply the \ac{idcc} approach to develop application programs for embedded systems on this scale.

\subsubsection{Observations}
During the development, the developers observed that organizing computations into separate functions significantly improved code quality and design.
This modular approach supports the \ac{idcc} approach since calls of such functions can be incorporated at an early stage of development when sketching the program's control flow.
The developers found that introducing variables separately often makes it easier to identify errors related to their use.

Interestingly, the developer of the \STM application program inspired its application development by an existing program (which, however, only supported part of the \Kimchi sensor chip's functionality).
The developer discovered that the existing program can be gradually decomposed to arrive at a program skeleton, forming the foundation for our \ac{idcc} approach.
This decomposition process can be viewed as applying the \ac{idcc} approach in reverse.

Throughout this work, both developers encountered certain bugs in the C~language support of \UltimateAutomizer.
Utilizing \ac{idcc} helped the developer in isolating these errors.
The developer was then able to replace the unsupported syntax of program statements with an alternative syntax that preserves the semantics.
Remarkably, our \ac{idcc} approach did not require any manual intervention by a developer, such as adding loop invariants to assist the software model checking tool~(\textbf{RQ3}).
\UltimateAutomizer was able to derive loop invariants and function contracts for each program revision automatically.

\subsubsection{Threads to Validity}
The concurrency introduced by interrupt handling is only simulated for the continuous checking.
Both developers do this by calling interrupt functions non-deterministically in the work loop of the program.
However, this does not capture all feasible executions of an interrupt-driven program.
A specialized software model checking tool for concurrent programs with support for interrupts could help to address this limitation.
In the two application programs, most statements including the \ac{hal} function calls are executed sequentially within the main program rather than in the interrupt context.
This is why the simulation of interrupt handling was not a major issue here.

Maintaining a verification harness during the development can be challenging.
The harness must include correct assumptions in order to detect temporal dependency violations.
Since the developers created the harness on-the-fly during the development they can maintain the harness in each development step.
This approach helps to reduce the risk of an incorrect harness compared to traditional testing where a harness with all its assumptions must be available before testing begins.

If the software model checking tool does not support certain features of the utilized programming language, continuous checking can fail during the application development.
This was the case in our evaluation but it could be traced back to syntactic issues.
Developers can address this by replacing problematic statements with equivalent ones that preserve semantics or by using a different model checker that supports the required language features.

Since there is no comparative application development to reference, it is hard to predict what the results would have been without applying the \ac{idcc} approach.
However, it is likely that various issues would emerge, as noted in our observations.
For instance, the absence of a verification harness could lead to wrong checking results, or the developer might deviate significantly from our suggested development guideline in \Cref{sec:incremental-development-guideline}.
Such deviations could cause significant changes to the program structure late in development, potentially causing the continuous checking to fail.

  % !TeX spellcheck = en_US

\section{Related Work}
\label{sec:related-work}
\ac{idcc} follows an incremental approach within the broader context of iterative and incremental methodologies outlined by Larman and Basili~\cite{Larman2003}.
Cockburn~\cite{Cockburn2008} highlights the differences of both methodologies.
The incremental approach establishes the expected functionality at the end of the development, whereas the iterative approach focuses on functionality first and then revise and improve its implementation until the end of the development.
Although the term \emph{incremental} of the \ac{idcc} approach suspect incremental \cite{Rothenberg2018,Seidl2020,Strichman2008} or witness-guided \cite{Saan2023} verification, \ac{idcc} does not require checking techniques, that reuse analysis results between program revisions.

In relation to existing software development approaches, \ac{idcc} establishes a \emph{verification-driven} approach that integrates quality assurance early in the application development, similar to test-driven development~\cite{Beck2002}.
Test-driven development extends the specification during development that the program must satisfy.
This is not the case within the \ac{idcc} approach since the specification is defined by the used \ac{hal} and available right from the start of the development.
The test cases that are created during the test-driven development are not meant to be used for software quality assurance.
The are only used to drive the development of the program forward without having to consider specific testing strategies with regard to quality assurance.
However, the test-driven development as well as \ac{idcc} contrast with traditional approaches where verification is typically considered as a separate and subsequent step.
\ac{idcc} itself is a subordinate workflow focused on the implementation and verification of software, especially during the \emph{program design}, \emph{coding}, and \emph{testing} steps in established development models such as the waterfall model~\cite{Benington1983,Royce1987} and the spiral model~\cite{Boehm1988}.
\ac{idcc} can also be integrated into approaches based on agile practices like Scrum~\cite{Schwaber2020}.

The incremental nature of \ac{idcc} is closely related with prototyping strategies~\cite{Crinnion1991,Luqi1989} where each program revision serves as a prototype.
Compared to model-driven development~\cite{Beydeda2005}, \ac{idcc} shares similarities with the work in \cite{Sirjani2021} in using incremental refinements although \ac{idcc} focuses on the development of application programs rather than just models.
\ac{idcc} also relates to design by contract~\cite{Meyer1991,Meyer1992} but it minimizes the burden on developers by not requiring an explicit specification of contracts.
Here, \ac{idcc} leverages the software model checking tool to automatically derive relevant pre- and postconditions.
There is no need to manually assists the software model checking tool, as seen in the creation of data abstractions in~\cite{Holzmann2004}.

Program development by stepwise refinement~\cite{Wirth1971,Back1990,Wehrheim2006,Ruhroth2012} emphasizes the importance of refining a program along its development and provides strategies for creating increments.
In contrast to our approach, stepwise refinement establishes a formal connection between program revisions to ensure that each new revision is a correct improvement over the last.
The creation of new revisions by increments can be based on established programming patterns from practice to improve the quality of the application design in the development of programs for embedded systems~\cite{Kim2009}.
The concepts of guarded commands and nondeterminacy~\cite{Dijkstra1975} further enhance the understanding of non-deterministic programming for the creation of initial program skeletons within \ac{idcc}.

Applying software model checking to application programs is well-established, as highlighted in surveys~\cite{Abdulla2004,Ball2011}.
A challenge in this area is overcoming the state space problem~\cite{Clarke2012}.
Nonetheless, several works have demonstrated that software model checking application programs succeeds.
This supports the motivation put forth by Reinbacher et al.~\cite{Reinbacher2008}.
An evaluation of software model checking tools applied to embedded C~programs is provided by~\cite{Schlich2009}.
Another work applied software model checking to application programs for embedded systems using the specification language PROMELA~\cite{Ribeiro2005}.
Several studies leverage software model checking to ensure the correct use of interfaces in application programs.
Specifically, the work by~\cite{Ball2006,Ball2004} employs the Static Driver Verifier from Microsoft to detect violations of the interface specification in Windows device drivers.
Similar efforts for Linux device drivers are reported in~\cite{Post2009}.
The case study in~\cite{Kolb2010} demonstrates how software model checking can check the proper usage of a communication stack \wrt a defined interface specification.
Compared to this work, software model checking is primarily applied at the end of application development to detect violations of the interface specification.
In contrast, \ac{idcc} focuses on enabling software model checking from the very beginning.

In this work, the specification for the temporal dependencies and the assumptions of the verification harness are encoded directly into the application code of each program revision.
This approach simplifies the use of software model checking tools, but specifications can alternatively be maintained outside the source code using different specification languages.
For instance, \textsc{SLIC} can be used for HAL interface specifications~\cite{Ball2001} while the verification harnesses can be implemented in languages presented in ~\cite{Groce2015,Ratiu2017}.

  % !TeX spellcheck = en_US

\section{Conclusion}
In this paper, we introduced \ac{idcc}, a novel approach for the incremental development of application programs for embedded systems that utilize \ac{hal} functions.
The focus of \ac{idcc} is on checking the \ac{hal} interface specification from the very beginning of the development.
The approach starts with a program skeleton, which is refined through a sequence of development steps to the final application program.
In each step, the continuous checking ensures that the \ac{hal} interface specification is satisfied.
We presented temporal dependencies as a \ac{hal} interface specification that allows continuous checking, even in a program skeleton with only the \ac{hal} function calls.
Our preliminary experimental evaluation demonstrates that \ac{idcc} is feasible in practice.
\ac{idcc} allows us to incrementally develop two application programs, each for an industrial embedded sensor system, where the continuous checking succeeds in each development step.

However, several open questions remain for future investigation.
Future work should include additional case studies within similar development contexts to assess the generalizability of \ac{idcc}.
It would be interesting to compare the results of these case studies with those from development projects using traditional approaches.
Moreover, exploring the applicability of \ac{idcc} with various software model checking tools is essential to evaluate its adaptability across different development environments.
Finally, investigating the inclusion of various use cases of automatically generated program skeletons from widely used \acp{ide} for embedded software development is crucial to improve the usability of \ac{idcc}.

  \begin{acks}
This work was partially funded by the \grantsponsor{gsp:dfg}{\acl{dfg} (\acs{dfg}, German Research Foundation)}{https://www.dfg.de/en} -- \grantnum{gsp:dfg}{503812980}.
\end{acks}

  \bibliographystyle{ACM-Reference-Format}
  \bibliography{content/bibliography}

@InCollection{Rothenberg2018,
  author    = {Bat-Chen Rothenberg and Daniel Dietsch and Matthias Heizmann},
  booktitle = {Static Analysis},
  publisher = {Springer International Publishing},
  title     = {Incremental Verification Using Trace Abstraction},
  year      = {2018},
  pages     = {364--382},
  doi       = {10.1007/978-3-319-99725-4_22},
  ranking   = {rank5},
}

@InCollection{Heizmann2018,
  author    = {Matthias Heizmann and Yu-Fang Chen and Daniel Dietsch and Marius Greitschus and Jochen Hoenicke and Yong Li and Alexander Nutz and Betim Musa and Christian Schilling and Tanja Schindler and Andreas Podelski},
  booktitle = {Tools and Algorithms for the Construction and Analysis of Systems},
  publisher = {Springer International Publishing},
  title     = {Ultimate Automizer and the Search for Perfect Interpolants},
  year      = {2018},
  pages     = {447--451},
  doi       = {10.1007/978-3-319-89963-3_30},
}

@InCollection{Heizmann2013a,
  author    = {Matthias Heizmann and Jochen Hoenicke and Andreas Podelski},
  booktitle = {Computer Aided Verification},
  publisher = {Springer Berlin Heidelberg},
  title     = {Software Model Checking for People Who Love Automata},
  year      = {2013},
  pages     = {36--52},
  doi       = {10.1007/978-3-642-39799-8_2},
}

@InCollection{Seidl2020,
  author    = {Helmut Seidl and Julian Erhard and Ralf Vogler},
  booktitle = {From Lambda Calculus to Cybersecurity Through Program Analysis},
  publisher = {Springer International Publishing},
  title     = {Incremental Abstract Interpretation},
  year      = {2020},
  pages     = {132--148},
  doi       = {10.1007/978-3-030-41103-9_5},
  ranking   = {rank4},
}

@Article{Ball2006,
  author    = {Ball, Thomas and Bounimova, Ella and Cook, Byron and Levin, Vladimir and Lichtenberg, Jakob and McGarvey, Con and Ondrusek, Bohus and Rajamani, Sriram K. and Ustuner, Abdullah},
  journal   = {ACM SIGOPS Operating Systems Review},
  title     = {Thorough static analysis of device drivers},
  year      = {2006},
  issn      = {0163-5980},
  month     = apr,
  number    = {4},
  pages     = {73--85},
  volume    = {40},
  doi       = {10.1145/1218063.1217943},
  publisher = {Association for Computing Machinery (ACM)},
}

@InBook{Ball2004,
  author    = {Ball, Thomas and Cook, Byron and Levin, Vladimir and Rajamani, Sriram K.},
  pages     = {1--20},
  publisher = {Springer Berlin Heidelberg},
  title     = {SLAM and Static Driver Verifier: Technology Transfer of Formal Methods inside Microsoft},
  year      = {2004},
  isbn      = {9783540247562},
  booktitle = {Integrated Formal Methods},
  doi       = {10.1007/978-3-540-24756-2_1},
  issn      = {1611-3349},
}

@InProceedings{Kolb2010,
  author    = {Kolb, Emanuel and Šerý, Ondřej and Weiss, Roland},
  booktitle = {Perspectives of Systems Informatics},
  title     = {Applicability of the {BLAST} Model Checker: An Industrial Case Study},
  year      = {2010},
  editor    = {Pnueli, Amir and Virbitskaite, Irina and Voronkov, Andrei},
  pages     = {218--229},
  publisher = {Springer},
  series    = {PSI},
  doi       = {10.1007/978-3-642-11486-1_19},
}

@Article{Post2009,
  author    = {Post, Hendrik and Sinz, Carsten and Küchlin, Wolfgang},
  journal   = {Software Testing, Verification and Reliability},
  title     = {Towards automatic software model checking of thousands of Linux modules -- a case study with Avinux},
  year      = {2009},
  number    = {2},
  pages     = {155--172},
  volume    = {19},
  doi       = {10.1002/stvr.399},
  publisher = {Wiley},
}

@Article{Ball2011,
  author    = {Ball, Thomas and Levin, Vladimir and Rajamani, Sriram K.},
  journal   = {Communications of the ACM},
  title     = {A decade of software model checking with SLAM},
  year      = {2011},
  issn      = {1557-7317},
  month     = jul,
  number    = {7},
  pages     = {68--76},
  volume    = {54},
  doi       = {10.1145/1965724.1965743},
  publisher = {Association for Computing Machinery (ACM)},
}

@InProceedings{Ball2001,
  author    = {Ball, Thomas and Rajamani, Sriram K.},
  booktitle = {Proceedings of the 8th International SPIN Workshop on Model Checking of Software},
  title     = {Automatically validating temporal safety properties of interfaces},
  year      = {2001},
  address   = {Berlin, Heidelberg},
  pages     = {103–122},
  publisher = {Springer-Verlag},
  series    = {SPIN '01},
  isbn      = {3540421246},
  location  = {Toronto, Ontario, Canada},
  numpages  = {20},
  url       = {https://dl.acm.org/doi/10.5555/380921.380932},
}

@Article{Wirth1971,
  author    = {Wirth, Niklaus},
  journal   = {Communications of the ACM},
  title     = {Program development by stepwise refinement},
  year      = {1971},
  issn      = {1557-7317},
  month     = apr,
  number    = {4},
  pages     = {221--227},
  volume    = {14},
  doi       = {10.1145/362575.362577},
  publisher = {Association for Computing Machinery (ACM)},
}

@Article{Boehm1988,
  author    = {Boehm, B. W.},
  journal   = {Computer},
  title     = {A spiral model of software development and enhancement},
  year      = {1988},
  issn      = {0018-9162},
  month     = may,
  number    = {5},
  pages     = {61--72},
  volume    = {21},
  doi       = {10.1109/2.59},
  publisher = {Institute of Electrical and Electronics Engineers (IEEE)},
}

@Article{Benington1983,
  author    = {Benington, Herbert D.},
  journal   = {IEEE Annals of the History of Computing},
  title     = {Production of Large Computer Programs},
  year      = {1983},
  issn      = {1058-6180},
  month     = oct,
  number    = {4},
  pages     = {350--361},
  volume    = {5},
  comment   = {Describes idea of the software development process, later called the waterfall model.},
  doi       = {10.1109/mahc.1983.10102},
  publisher = {Institute of Electrical and Electronics Engineers (IEEE)},
}

@InProceedings{Royce1987,
  author    = {Royce, W. W.},
  booktitle = {Proceedings of the 9th International Conference on Software Engineering},
  title     = {Managing the development of large software systems: concepts and techniques},
  year      = {1987},
  address   = {Washington, DC, USA},
  pages     = {328--338},
  publisher = {IEEE Computer Society Press},
  series    = {ICSE '87},
  comment   = {Describes the water fall model with its process steps.},
  isbn      = {0897912160},
  location  = {Monterey, California, USA},
  numpages  = {11},
}

@Article{Luqi1989,
  author    = {Luqi},
  journal   = {Computer},
  title     = {Software evolution through rapid prototyping},
  year      = {1989},
  issn      = {0018-9162},
  month     = may,
  number    = {5},
  pages     = {13--25},
  volume    = {22},
  doi       = {10.1109/2.27953},
  publisher = {Institute of Electrical and Electronics Engineers (IEEE)},
}

@Book{Crinnion1991,
  author    = {Crinnion, John},
  editor    = {DeMIllo, Richard A.},
  publisher = {Plenum Press},
  title     = {Evolutionary Systems Development: A Practical Guide to the Use of Prototyping Within a Structured Systems Methodology},
  year      = {1991},
  address   = {New York, US},
  isbn      = {9780306441394},
  series    = {Software science and engineering},
}

@Article{Larman2003,
  author    = {Larman, C. and Basili, V.R.},
  journal   = {Computer},
  title     = {Iterative and Incremental Development: A Brief History},
  year      = {2003},
  issn      = {0018-9162},
  month     = jun,
  number    = {6},
  pages     = {47--56},
  volume    = {36},
  doi       = {10.1109/mc.2003.1204375},
  publisher = {Institute of Electrical and Electronics Engineers (IEEE)},
}

@Article{Cockburn2008,
  author   = {Alistair Cockburn},
  journal  = {STSC CrossTalk},
  title    = {Using Both Incremental and Iterative Development},
  year     = {2008},
  month    = may,
  number   = {5},
  pages    = {27--30},
  volume   = {21},
  url      = {https://web.archive.org/web/20120526153630/http://www.crosstalkonline.org/storage/issue-archives/2008/200805/200805-Cockburn.pdf},
}

@Article{Meyer1992,
  author    = {Meyer, Bertrand},
  journal   = {Computer},
  title     = {Applying “design by contract”},
  year      = {1992},
  issn      = {0018-9162},
  month     = oct,
  number    = {10},
  pages     = {40--51},
  volume    = {25},
  doi       = {10.1109/2.161279},
  publisher = {Institute of Electrical and Electronics Engineers (IEEE)},
}

@InBook{Meyer1991,
  author    = {Meyer, Bertrand},
  editor    = {Mandrioli, Dino and Meyer, Bertrand},
  pages     = {1--50},
  publisher = {Prentice Hall},
  title     = {Design by Contract},
  year      = {1991},
  address   = {New York, London},
  isbn      = {978-0130065780},
  booktitle = {Advances in Object-Oriented Software Engineering},
  url       = {https://se.inf.ethz.ch/~meyer/publications/old/dbc_chapter.pdf},
}

@InBook{Strichman2008,
  author    = {Strichman, Ofer and Godlin, Benny},
  pages     = {496--501},
  publisher = {Springer Berlin Heidelberg},
  title     = {Regression Verification - A Practical Way to Verify Programs},
  year      = {2008},
  isbn      = {9783540691495},
  booktitle = {Verified Software: Theories, Tools, Experiments},
  doi       = {10.1007/978-3-540-69149-5_54},
  issn      = {1611-3349},
}

@InBook{Saan2023,
  author    = {Saan, Simmo and Schwarz, Michael and Erhard, Julian and Seidl, Helmut and Tilscher, Sarah and Vojdani, Vesal},
  pages     = {74--97},
  publisher = {Springer Nature Switzerland},
  title     = {Correctness Witness Validation by Abstract Interpretation},
  year      = {2023},
  isbn      = {9783031505249},
  month     = dec,
  booktitle = {Verification, Model Checking, and Abstract Interpretation},
  doi       = {10.1007/978-3-031-50524-9_4},
  issn      = {1611-3349},
}

@Book{Beck2002,
  author    = {Kent Beck},
  publisher = {Pearson International},
  title     = {Test Driven Development: By Example},
  year      = {2002},
  edition   = {1},
  isbn      = {978-0-321-14653-3},
  month     = nov,
}

@Article{Sirjani2021,
  author    = {Sirjani, Marjan and Provenzano, Luciana and Asadollah, Sara Abbaspour and Moghadam, Mahshid Helali and Saadatmand, Mehrdad},
  journal   = {Journal of Internet Services and Applications},
  title     = {Towards a Verification-Driven Iterative Development of Software for Safety-Critical Cyber-Physical Systems},
  year      = {2021},
  issn      = {1869-0238},
  month     = may,
  number    = {1},
  volume    = {12},
  doi       = {10.1186/s13174-021-00132-z},
  publisher = {Sociedade Brasileira de Computacao - SB},
}

@Book{Beydeda2005,
  editor    = {Sami Beydeda, Matthias Book, Volker Gruhn},
  publisher = {Springer Berlin Heidelberg},
  title     = {Model-Driven Software Development},
  year      = {2005},
  isbn      = {9783540285540},
  doi       = {10.1007/3-540-28554-7},
}

@Article{Dijkstra1975,
  author    = {Dijkstra, Edsger W.},
  journal   = {Communications of the ACM},
  title     = {Guarded commands, nondeterminacy and formal derivation of programs},
  year      = {1975},
  issn      = {1557-7317},
  month     = aug,
  number    = {8},
  pages     = {453--457},
  volume    = {18},
  doi       = {10.1145/360933.360975},
  publisher = {Association for Computing Machinery (ACM)},
}

@Misc{Schwaber2020,
  author  = {Ken Schwaber and Jeff Sutherland},
  month   = nov,
  title   = {The Scrum Guide --- The Definitive Guide to Scrum: The Rules of the Game},
  year    = {2020},
  url     = {https://scrumguides.org/docs/scrumguide/v2020/2020-Scrum-Guide-US.pdf},
  urldate = {2024-10-13},
}

@InBook{Clarke2012,
  author    = {Clarke, Edmund M. and Klieber, William and Nováček, Miloš and Zuliani, Paolo},
  pages     = {1--30},
  publisher = {Springer Berlin Heidelberg},
  title     = {Model Checking and the State Explosion Problem},
  year      = {2012},
  isbn      = {9783642357466},
  booktitle = {Tools for Practical Software Verification},
  doi       = {10.1007/978-3-642-35746-6_1},
  issn      = {1611-3349},
}

@InBook{Abdulla2004,
  author    = {Abdulla, Parosh Aziz and Jonsson, Bengt and Nilsson, Marcus and Saksena, Mayank},
  pages     = {35--48},
  publisher = {Springer Berlin Heidelberg},
  title     = {A Survey of Regular Model Checking},
  year      = {2004},
  isbn      = {9783540286448},
  booktitle = {CONCUR 2004 - Concurrency Theory},
  doi       = {10.1007/978-3-540-28644-8_3},
  issn      = {1611-3349},
}

@InProceedings{Beyer2024a,
  author    = {Beyer, Dirk},
  booktitle = {Tools and Algorithms for the Construction and Analysis of Systems},
  title     = {State of the Art in Software Verification and Witness Validation: SV-COMP 2024},
  year      = {2024},
  address   = {Cham},
  editor    = {Finkbeiner, Bernd and Kov{\'a}cs, Laura},
  pages     = {299--329},
  publisher = {Springer Nature Switzerland},
  doi       = {10.1007/978-3-031-57256-2_15},
  isbn      = {978-3-031-57256-2},
}

@InProceedings{Kim2009,
  author    = {Kim, Doo-Hwan and Kim, Jong-Phil and Hong, Jang-Eui},
  booktitle = {2009 Ninth International Conference on Quality Software},
  title     = {Practice Patterns to Improve the Quality of Design Model in Embedded Software Development},
  year      = {2009},
  month     = aug,
  pages     = {179--184},
  publisher = {IEEE},
  doi       = {10.1109/qsic.2009.32},
}

@InProceedings{Ratiu2017,
  author     = {Ratiu, Daniel and Ulrich, Andreas},
  booktitle  = {Proceedings of the 24th ACM SIGSOFT International SPIN Symposium on Model Checking of Software},
  title      = {Increasing usability of spin-based C code verification using a harness definition language: leveraging model-driven code checking to practitioners},
  year       = {2017},
  month      = jul,
  pages      = {60--69},
  publisher  = {ACM},
  series     = {ISSTA ’17},
  collection = {ISSTA ’17},
  doi        = {10.1145/3092282.3092283},
}

@InBook{Groce2015,
  author    = {Groce, Alex and Pinto, Jervis},
  pages     = {204--218},
  publisher = {Springer International Publishing},
  title     = {A Little Language for Testing},
  year      = {2015},
  isbn      = {9783319175249},
  booktitle = {NASA Formal Methods},
  doi       = {10.1007/978-3-319-17524-9_15},
  issn      = {1611-3349},
}

@InBook{Holzmann2004,
  author    = {Holzmann, Gerard J. and Joshi, Rajeev},
  pages     = {76--91},
  publisher = {Springer Berlin Heidelberg},
  title     = {Model-Driven Software Verification},
  year      = {2004},
  isbn      = {9783540247326},
  booktitle = {Model Checking Software},
  doi       = {10.1007/978-3-540-24732-6_6},
  issn      = {1611-3349},
}

@InProceedings{Reinbacher2008,
  author    = {Reinbacher, Thomas and Kramer, Michael and Horauer, Martin and Schlich, Bastian},
  booktitle = {2008 IEEE/ASME International Conference on Mechtronic and Embedded Systems and Applications},
  title     = {Motivating Model Checking of Embedded Systems Software},
  year      = {2008},
  month     = oct,
  pages     = {546--551},
  publisher = {IEEE},
  doi       = {10.1109/mesa.2008.4735653},
}

@InProceedings{Ribeiro2005,
  author    = {Ribeiro, O.R. and Fernandes, J.M. and Pinto, L.F.},
  booktitle = {12th IEEE International Conference and Workshops on the Engineering of Computer-Based Systems (ECBS’05)},
  title     = {Model Checking Embedded Systems with PROMELA},
  year      = {2005},
  pages     = {378--385},
  publisher = {IEEE},
  doi       = {10.1109/ecbs.2005.53},
}

@Article{Schlich2009,
  author    = {Schlich, Bastian and Kowalewski, Stefan},
  journal   = {International Journal on Software Tools for Technology Transfer},
  title     = {Model checking C source code for embedded systems},
  year      = {2009},
  issn      = {1433-2787},
  month     = mar,
  number    = {3},
  pages     = {187--202},
  volume    = {11},
  doi       = {10.1007/s10009-009-0106-5},
  publisher = {Springer Science and Business Media LLC},
}

@Article{Kuhl2013,
  author    = {Kuhl, Matthias and Gieschke, Pascal and Rossbach, Daniel and Hilzensauer, Sascha A. and Panchaphongsaphak, Thanapon and Ruther, Patrick and Lapatki, Bernd G. and Paul, Oliver and Manoli, Yiannos},
  journal   = {IEEE Journal of Solid-State Circuits},
  title     = {A Wireless Stress Mapping System for Orthodontic Brackets Using CMOS Integrated Sensors},
  year      = {2013},
  issn      = {1558-173X},
  month     = sep,
  number    = {9},
  pages     = {2191--2202},
  volume    = {48},
  doi       = {10.1109/jssc.2013.2264619},
  publisher = {Institute of Electrical and Electronics Engineers (IEEE)},
}

@InProceedings{Allinger2023,
  author    = {Allinger, Kim and Kuhl, Matthias},
  booktitle = {International Solid-State Circuits Conference},
  title     = {A Closed-Loop 12bit {CMOS}-Integrated Stress Sensor System with 4bit Adjustable Sensitivity from 178 to 11 kPa/LSB at up to 22.5kS/s and 5bit Dynamic Range Adjustment},
  year      = {2023},
  month     = feb,
  pages     = {1--3},
  publisher = {IEEE},
  series    = {ISSCC},
  doi       = {10.1109/ISSCC42615.2023.10067788},
}

@Book{ISO2011C,
  author    = {{ISO}},
  publisher = {International Organization for Standardization},
  title     = {{ISO}/{IEC} 9899:2011 Information technology --- Programming languages --- {C}},
  year      = {2011},
  address   = {Geneva, Switzerland},
  edition   = {3},
  month     = dec,
  bibdate   = {Mon Dec 19 11:12:12 2011},
  day       = {8},
  pages     = {683},
  remark    = {Revises ISO/IEC 9899:1999, ISO/IEC 9899:1999/Cor 1:2001, ISO/IEC 9899:1999/Cor 2:2004, and ISO/IEC 9899:1999/Cor 3:2007.},
  url       = {https://www.iso.org/standard/57853.html},
}

@Article{Ruhroth2012,
  author    = {Ruhroth, Thomas and Wehrheim, Heike},
  journal   = {Science of Computer Programming},
  title     = {Model evolution and refinement},
  year      = {2012},
  issn      = {0167-6423},
  month     = mar,
  number    = {3},
  pages     = {270--289},
  volume    = {77},
  doi       = {10.1016/j.scico.2011.04.007},
  publisher = {Elsevier BV},
}

@InProceedings{Wehrheim2006,
  author    = {Wehrheim, Heike},
  booktitle = {Architecting Systems with Trustworthy Components},
  title     = {Refinement and Consistency in Component Models with Multiple Views},
  year      = {2006},
  pages     = {84--102},
  publisher = {Springer Berlin Heidelberg},
  doi       = {10.1007/11786160_5},
  isbn      = {9783540358336},
  issn      = {1611-3349},
}

@InProceedings{Back1990,
  author    = {Back, R. J. R. and Wright, J.},
  booktitle = {Stepwise Refinement of Distributed Systems Models, Formalisms, Correctness},
  title     = {Refinement calculus, part I: Sequential nondeterministic programs},
  year      = {1990},
  pages     = {42--66},
  publisher = {Springer Berlin Heidelberg},
  series    = {REX},
  doi       = {10.1007/3-540-52559-9_60},
  isbn      = {9783540470359},
  issn      = {1611-3349},
}

\end{document}